\documentclass[aps,twocolumn,prc,amssymb,amsmath,bm,floatfix,superscriptaddress,nofootinbib,float]{revtex4}
\usepackage[dvipsnames,usenames]{xcolor}
\usepackage{graphicx}
\usepackage{mathptmx}
\usepackage{bm}

\newcommand{\be}{\begin{equation}}
\newcommand{\ee}{\end{equation}}
\newcommand{\bea}{\begin{eqnarray}}
\newcommand{\eea}{\end{eqnarray}}
\newcommand{\ba}{\begin{align}}
\newcommand{\ea}{\end{align}}

\newcommand{\para}{\parallel}

\usepackage{setspace}

\begin{document}

\title{Elastic properties of phases with nonspherical nuclei in dense matter }

\author{C. J. Pethick}
\affiliation{The Niels Bohr International Academy, The Niels Bohr Institute, University of Copenhagen, Blegdamsvej 17, DK-2100 Copenhagen \O, Denmark}
\affiliation{NORDITA, KTH Royal Institute of Technology and Stockholm University, Roslagstullsbacken 23, SE-106 91 Stockholm, Sweden}
\author{Zhaowen Zhang}
\affiliation{The Niels Bohr International Academy, The Niels Bohr Institute, University of Copenhagen, Blegdamsvej 17, DK-2100 Copenhagen \O, Denmark}
\author{D. N. Kobyakov}
\affiliation{Institute of Applied Physics of the Russian Academy of Sciences, 603950 Nizhny Novgorod, Russia}
\begin{abstract}
We consider the elastic constants of phases with nonspherical nuclei, so-called pasta phases, predicted to occur in the inner crust of a neutron star.  First, we treat perfectly ordered phases and give numerical estimates for lasagna and spaghetti when the pasta elements are spatially uniform: the results are in order-of-magnitude agreement with the numerical simulations of Caplan, Schneider, and Horowitz, Phys.\ Rev.\ Lett.\ {\bf 121}, 132701 (2018).   We then turn to pasta phases without long-range order and calculate upper (Voigt) and lower (Reuss) bounds on the effective shear modulus and find that the lower bound is zero, but the upper bound is nonzero.  To obtain better estimates, we then apply the self-consistent formalism and find that this predicts that the shear modulus of the phases without long-range order is zero if the pasta elements are spatially uniform.  In numerical simulations, the pasta elements are found to be modulated spatially and we show that this modulation is crucial to obtaining a nonzero elastic moduli for pasta phases without long-range order.      In the self-consistent formalism we find that, for lasagna,  the effective shear modulus is linear in the elastic constants that do not vanish when the pasta elements are spatially uniform while, for spaghetti, it varies as the square root of these elastic constants.  We also consider the behavior of the elastic constant associated with a homologous strain (hydrostatic compression) of the structure of the pasta phases without long-range order.
\end{abstract}

\maketitle

\section{Introduction}
Elastic properties of crusts of neutron stars are important for modelling stellar oscillations and for estimating continuous gravitational wave emission from rotating neutron stars \cite{ Abbott}. Theoretical considerations strongly suggest that at densities just below that of nuclear matter,  nuclei may be rod-like or plate-like, rather than roughly spherical \cite{pasta, CJPRavenhall}.  Such states are referred to as ``pasta'' phases because of their resemblance to spaghetti and lasagna.\footnote{For brevity, we shall refer to the phase with rod-like nuclei as ``spaghetti'' and the phase with plate-like nuclei as ``lasagna''.}   For matter in stellar collapse, in which the proton fraction is relatively high, $\sim 0.3$, the pasta phases are robust, in the sense that their appearance  is relatively insensitive to details of the nuclear Hamiltonian and the many-body methods employed \cite{WilliamsKoonin, NewtonStone,PaisStone,Sonoda,Horowitz}.    For the lower proton fractions encountered in neutron stars at densities close to nuclear density the calculations of Ref.~\cite{Lorenz, OyamatsuNS} indicate that pasta phases are the ground state.  However, the appearance of these phases does depend on the nuclear Hamiltonian: for the SLy4 nuclear interaction Douchin and Haensel \cite{DouchinHaensel} found that, with increasing density, matter underwent a transition from round nuclei to a uniform liquid without passing through the pasta phases. In a parameter study of a family of nuclear Hamiltonians based on relativistic mean field theory, Bao and Shen showed that the appearance of pasta phases was correlated with the size of the parameter $L$, the derivative of the symmetry energy with respect to the logarithm of the density \cite{BaoShen}.  The pasta phases could constitute a large fraction of the mass of the crust of a neutron star, but just how large depends on details of the nuclear Hamiltonian and further work is needed to clarify the issue. 

The elastic properties of the lasagna phase have recently been calculated in  molecular dynamics simulations \cite{Caplan}, and
the purpose of this article is to perform analytical calculations of  elastic properties of the pasta phases. Building on the work of  Ref.~\cite{CJPPotekhin}, we begin by considering elastic constants of ordered phases.  However, it is to be expected that the pasta structures in neutron stars will not be uniformly \mbox{oriented}.  For random orientations of the pasta, the elastic properties on length scales large compared with the characteristic length scale for variations of the orientation of the pasta may be described by those of an isotropic medium.  The system may be characterized in terms of the bulk and shear strains of the periodic structure of the pasta elements, in addition to the densities of neutrons and protons.   The main part of the present article is devoted to calculating effective elastic constants $\mu$ for a shear strain and $K^u$ for a strain without shear.  For conventional solids the latter strain is referred to as a hydrostatic compression in the language of Ref.\ \cite{LandLElasticity} but, for the pasta phases, it is only the structure of the phases which is strained, and densities are held constant.  

Study of problems of the elastic properties of polycrystalline matter has a long history in materials science and geophysics, and two of the  earliest approximations are those of Voigt \cite{Voigt},  who assumed that the local strain in the medium is constant everywhere, and Reuss \cite{Reuss},  who assumed that the local stress in the medium is constant everywhere. Hill demonstrated that the Voigt approximation gives an upper bound on the effective elastic constants and the Reuss approximation a lower bound \cite{Hill}.  As Berryman has reviewed in detail \cite{Berryman2005}, subsequent work proceeded in two directions:  the derivation of improved upper and lower bounds \cite{Kube_deJong} and the development of the so-called self-consistent approach.   These methods have been very successful in accounting for the properties of terrestrial materials and have previously been applied to astrophysical solids with cubic symmetry \cite{KobyakovCJP2015}. 

The pasta phases are much more anisotropic than most terrestrial materials, and we shall show that for pasta with {\it uniform} spaghetti strands or lasagna sheets, the Reuss lower bounds and the results of the self-consistent method are all zero, while the Voigt bounds are nonzero.   Numerical simulations to date indicate that the pasta elements are generally modulated  spatially in directions in the plane of the lasagna sheets or the direction of the spaghetti strands  \cite{WilliamsKoonin, NewtonStone,PaisStone,Sonoda,Horowitz}, and we show that when this is taken into account, the predicted effective elastic constants are nonzero.

The plan of the paper is that in Sec.\ II we describe the basic formalism and in Sec.\ III we give numerical estimates for the elastic constants of lasagna and spaghetti without spatial modulation.  Section IV presents the Voigt and Reuss bounds on the effective shear and bulk moduli for ``polycrystalline'' pasta phases, as well as the self-consistent formalism, while Sec.\ V is a brief concluding section.

\section{Basic considerations}

In this section we first describe the elastic properties of perfect lasagna and spaghetti structures, i.e., ones with long range order.    We shall assume that the perfectly ordered pasta phases have hexagonal symmetry.  This is consistent with what  molecular dynamics  simulations indicate for lasagna \cite{SchneiderWaffles} and spaghetti \cite{Sonoda}.\footnote{Numerical Thomas--Fermi or Hartree--Fock calculations use much smaller, cubic computational cells, and the crystal structure is then strongly influenced by the shape of the cell.}\textsuperscript{,}\footnote{In Ref.\ \cite{SchneiderWaffles} phases with spatially modulated plate-like elements are referred to as ``waffles'' but, to avoid proliferation of nomenclature, we shall refer to them as ``lasagna''. } In the body of the paper we shall not take magnetic fields into account, but we consider them briefly in Sec.\ \ref{summary}.  For a hexagonal crystal, the second order elastic properties are invariant under rotations about the $z$-axis and the general form for the elastic energy per unit volume is \cite[\S10]{LandLElasticity}
\begin{multline}
E_{\rm elast}=\frac12 c_{11}(u_{xx}+u_{yy})^2  +\frac12 c_{33}u_{zz}^2 +c_{13}u_{zz}(u_{xx}+u_{yy})\\
 +  2c_{44}(u_{xz}^2+u_{yz}^2)  +2c_{66}( u_{xy}^2 -u_{xx}u_{yy} )
 \end{multline}
\begin{multline}
 =\frac12 c_{33}u_{zz}^2 +\frac12 \frac{(c_{11}+c_{12})}{2}(u_{xx}+u_{yy})^2 \\+\frac12 c_{66}\left( (u_{xx}-u_{yy})^2+4 u_{xy}^2\right)\ \\  +c_{13}u_{zz}(u_{xx}+u_{yy})]
 +  2c_{44}(u_{xz}^2+u_{yz}^2),
 \label{ElasticEnergy}
\end{multline}
where the strain tensor is given to first order in the displacement vector $\bf u$ by\footnote{We shall work in terms of the tensor strains defined by the equation that follows.  Some authors work in terms of so-called ``engineering strains'' $e_{ij}$, where $e_{ij}=u_{ij}$ for $i=j$ but  $e_{ij}=2u_{ij}$ for $i\neq j$, see, e.g., Ref. \cite{Berryman2005}.}
\be
u_{ij}=\frac12\left( \frac{\partial u_i}{\partial x_j} +\frac{\partial u_j}{\partial x_i}  \right).
\ee

For brevity and clarity we use the Voigt notation ($1={xx},~ 2={yy},~3={zz},~4={yz},~ 5={zx},$ and $6={xy}$).  Because of the rotational symmetry about the $z$-axis, $c_{66}=(c_{11}-c_{12})/2$. The elastic constants $c_{ij}$ are related to those defined in Ref. \cite{LandLElasticity} by $c_{33}=\lambda_{zzzz}$, $c_{11}=\lambda_{xxxx}$, $c_{66}=\lambda_{xyxy}$, 
$c_{13}=\lambda_{xxzz}$, and $c_{44}=\lambda_{xzxz}=\lambda_{yzyz}$.  

The energy density contains, in addition,  terms involving changes in the densities of neutrons and protons and ones that involve both density changes and strains.  We shall consider the case of disturbances with wavelengths long compared with the electron screening length, so the coarse-grained average density of electrons is equal to that of the protons, thus ensuring that matter is electrically neutral.  Consequently, there are only two independent densities, that of the neutrons, and that of the charged particles.  
To second order in the density changes, they contribute to the energy density an amount
\be
E_{\rm dens}=\frac12 E_{\alpha\beta}\delta n_\alpha \delta n_\beta,
\label{Edens}
\ee
where 
\be
E_{\alpha\beta}=\frac{\partial^2 E}{\partial n_\alpha \partial n_\beta}.
\ee
Here $\alpha$ and $\beta$ label the species, $n$ for neutron and $c$ for the charged particles, and we employ the summation convention for indices.
The terms in the energy density that couple  strains and density variations are given in lowest order by  \cite{KobyakovCJP2018,DurelUrban,KobyakovCJP2020}
\be
E_{\rm coup}= \left[C_{\alpha\para}u_{zz}+C_{\alpha\perp}(u_{xx}+u_{yy}) \right]\delta n_\alpha,
\label{Ecoup}
\ee 
where
\be
C_{\alpha\para}=\frac{\partial^2 E}{\partial n_\alpha \partial u_{zz}}\;\;\;{\rm and}\;\;\;C_{\alpha\perp}=\frac{\partial^2 E}{\partial n_\alpha \partial u_{xx}}.  
\ee
In the coupling energy, terms linear in $u_{ij}$ with $i \neq j$ cannot occur due to the requirement of rotational invariance. 

For the pasta phases and for strains and relative changes of the densities of order unity, the elastic  and coupling energy densities are of order the surface and Coulomb energy densities, while the energy density due to density changes is of the order of typical bulk energy densities, which are considerably larger than the surface and Coulomb energy densities. 

It is important to note that the elastic constants $c_{ij}$ defined above are related to changes in the energy of the system {\it for fixed particle densities}.  For conventional solids with isolated nuclei, the densities of nucleons and electrons are constrained, since the only way in which nucleon densities can change is by displacement of nuclei, and consequently for such systems, the elastic constants will contain contributions due to density changes; in particular, $c_{ij}$ for $i,j\leq 3$ are of the order of bulk energy densities, not surface and Coulomb ones.   Here we shall consider only static situations and, consequently, we shall not consider kinetic contributions to the energy density, and viscous stresses. 

\section{Elastic constants of lasagna and spaghetti phases}

For lasagna and spaghetti without modulations, many of the elastic constants vanish: for lasagna only $c_{33}$ is nonzero and, for spaghetti, only $c_{11} (=c_{22})$, $c_{12}$ and $c_{66}=(c_{11}-c_{12})/2$ are.  These elastic constants have been calculated in Ref.\ \cite{CJPPotekhin} within the framework of a liquid drop picture.\footnote{In Ref.\ \cite{CJPPotekhin},  for lasagna, $c_{33}$ is denoted by $B$ and, for spaghetti, \mbox{$(c_{11}+c_{12})/2$} by $B$ and $(c_{11}-c_{12})/2$ by $C$.}
For lasagna the result is
\be
c_{33}=3 E_{\rm surf},
\label{c33}
\ee
where $E_{\rm surf}$ is the surface energy per unit volume, while for spaghetti one finds
\be
\frac{c_{11}+c_{12}}{2}=\frac{3}{4}E_{\rm surf}
\label{c11+c12}
\ee 
and
\be
c_{66}=\frac{c_{11}-c_{12}}{2}=E_{\rm Coul} \,g(u)=\frac{E_{\rm surf}}{2} \,g(u).
\label{c66}
\ee
For the range of values of $u$ for which the spaghetti phase is expected to be stable, the function $g(u)$ is well fitted by the expression
\be
g(u)=10^{2.1(u-0.3)}.
\ee 
 The surface energy per unit volume for the two phases is given by
\be
E_{\rm surf}=\frac{\sigma u d}{r_N},
\ee
where $d$ is the dimensionality of the structure (1 for lasagna, 2 for spaghetti, and 3 for spherical nuclei)  and $u=(r_N/r_c)^d$ is the fraction of space occupied by nuclear matter.  Half the thickness of a lasagna sheet, the radius of a spaghetti strand and the radius of a spherical nucleus are denoted by $r_N$, and the quantity $r_c$ is the radius of a sphere with volume equal to the average volume per spherical nucleus, the radius of a circle with area equal to the average area per spaghetti strand, and half the spacing of lasagna sheets.  
The Coulomb energy per unit volume is given by 
\be
E_{\rm Coul}=2\pi (n_{pi}e r_N)^2 uf_d(u),
\ee
where $n_{pi}$ is the proton density within a lasagna sheet or a spaghetti strand, and
\be
f_d(u)=\frac{1}{d+2}\left[ \frac{2}{d-2}\left(1-\frac{du^{1-2/d}}{2}\right) +u   \right].
\ee
  Specifically, 
\be
f_1(u)=\frac13 \frac{(1-u)^2}{u},
\ee
and 
\be
f_2(u)=\frac{1}{4}\left(\ln \frac{1}{u}-1+u\right).
\ee
The total Coulomb and surface energy is a minimum for given filling factor $u$ when
\be
E_{\rm surf}=2E_{\rm Coul},
\ee
and therefore
\be
r_{N}=\left(\frac{d \sigma}{4 \pi (n_{pi} e)^2 f_d(u)}\right)^{1/3}.
\label{rN}
\ee
For lasagna, one finds from Eqs.\ (\ref{c33}) and (\ref{rN}) that
\begin{align}
c_{33}&=(36 \pi)^{1/3} [n_{pi}\, e\,\sigma\, u(1-u)]^{2/3} \\
&\approx 1.61 \left(\frac{n_{pi}}{n_s}\frac{\sigma}{1 {\rm MeV/fm}^2} u(1-u)\right)^{2/3} \,  {\rm MeV\, fm}^{-3}\\
&\approx 2.58 \cdot 10^{33}  \left(\frac{n_{pi}}{n_s}\frac{\sigma}{1 {\rm MeV/fm}^2} u(1-u)\right)^{2/3} \,  {\rm erg\, cm}^{-3}.
\end{align}
To obtain an order of magnitude estimate, we take $\sigma=0.1$ MeV\, fm$^{-2}$, $n_{pi}\approx 0.05 n_s$ \cite{OyamatsuNS} and $u=0.5$.  One then finds $c_{33}\approx  3\cdot 10^{31}$ erg cm$^{-3}$.
For spaghetti, the analogous results are
\begin{align}
\frac12(c_{11}+c_{12})&=  \frac32 (2 \pi )^{1/3}(n_{pi}\, e\,\sigma)^{2/3}  u\left[f_2(u)\right]^{1/3} \\
&\approx  0.92\left(\frac{n_{pi}}{n_s}\frac{\sigma}{1 {\rm MeV/fm}^2}\right)^{2/3}u\left[f_2(u)\right]^{1/3} \,  {\rm MeV\, fm}^{-3}\\
&\approx 1.48 \cdot 10^{33} \left(\frac{n_{pi}}{n_s}\frac{\sigma}{1 {\rm MeV/fm}^2}\right)^{2/3}u\left[f_2(u)\right]^{1/3} \,  {\rm erg\, cm}^{-3}
\end{align}
and
\be
c_{66}=\frac43 \frac{(c_{11}+c_{12})}2 g_2(u) .
\ee
For $\sigma=0.1$ MeV\, fm$^{-2}$, $n_{pi}=0.1 n_s$ \cite{OyamatsuNS} and $u=0.3$,   one finds $(c_{11}+c_{12})/2\approx 1\cdot 10^{31}$ erg cm$^{-3}$, and $c_{66}$ is about one half of this.

The above results bring out clearly the importance of the surface tension of nuclear matter. The orders of magnitude of the elastic constants we find are similar to those found in the simulations of Caplan et al.\ \cite{Caplan} for proton fractions \mbox{$\sim 0.3$.}  Detailed comparison is not possible because the surface energy for the nucleon--nucleon interaction used in that work has not been evaluated.

As mentioned in the Introduction, numerical simulations of the pasta phases indicate that the pasta elements are not uniform, but are corrugated.  In general, all five elastic constants in Eq.\ (\ref{ElasticEnergy}) are nonzero.  As we shall show, this has important implications for the elastic properties of  ``polycrystalline'' pasta.

\section{``Polycrystalline'' Pasta}

In astrophysical environments, it is unlikely that the pasta phases are well ordered throughout a star.  Rather one expects the symmetry axes of the pasta to vary  with position.  Exactly what form the spatial variation of the axes takes is unknown.  One possibility is that there are well defined domains with constant directions of the symmetry axes, with abrupt changes of direction from one domain to another, in much the same way as in a polycrystalline solid.  Another possibility is that the directions of the symmetry axes vary smoothly in space, in a manner similar to what is observed in laboratory liquid crystals \cite{deGennesProst}.  We shall use the word ``polycrystalline'' to describe both situations. In our discussion, we shall assume that, locally, the material has a well-defined periodic structure, that variations of the axes of the structure vary only on length scales large compared with lattice spacings, and that there are no long-range correlations between the symmetry directions of the local structure.   The basic assumption made in the work that follows is that the total elastic energy is given by the volume integral of the local energy density, Eq.\ (\ref{ElasticEnergy}).  Thus contributions to the energy from domain walls or the distortions of the pasta structure from a uniform phase are neglected.  This should be a good approximation provided the domains have a size much larger than the lattice spacing or that the variations of the direction of the symmetry axes occur on length scales large compared with the lattice spacing.     In what follows, we shall use the language of polycrystals but the results also apply to the case of continuous variations of the directions of the symmetry axes.  

To second order in deviations from an initial uniform state, the general expression for the energy density of polycrystalline phases is given by
\be
E= \mu \left(u_{ij} -  \frac13 u_{ll}\delta_{ij}  \right)^2  +\frac12 K^u u_{ll}^2  +         \frac12 E_{\alpha\beta}\delta n_\alpha \delta n_\beta + C_\alpha u_{ii}\delta n_\alpha.
\label{EPoly}
\ee  
Here $\mu$ is an effective shear elastic constant and $K^u$ is an effective bulk modulus associated with straining the lasagna structure, keeping the densities of the components fixed. 
The strains and density changes are now considered to be coarse-grained averages but, for simplicity, we shall not indicate this explicitly.  Since the medium is effectively isotropic, the coefficients $C_\alpha$ are independent of the direction of the diagonal strains.
The last two terms in Eq.\ (\ref{EPoly}) are analogous to the contributions (\ref{Edens}) and (\ref{Ecoup}) for pasta phases with long-range order.
In a usual solid with isolated nuclei, the neutron and proton density changes are related directly to the displacements of the nuclei, and therefore the energy density may be written in terms of  the $u_{ij}$ alone.

When a polycrystal is subjected to a strain, differences in chemical potentials between different crystallites will arise because chemical potentials depend on strains as a consequence of the contributions to the energy in Eq.\ (\ref{Ecoup}).  In equilibrium, the chemical potentials of neutrons and charged particles will be constant throughout the polycrystal, but because the strains within the various crystallites are not the same, the densities of particles with not be constant.  This implies that the elastic constants of single crystals that are relevant for determining the effective elastic constants  of polycrystals are related to energy changes \textit{for fixed chemical potentials}.  However, as described in    \cite[\S 45]{LandLElasticity} the elastic constants at constant chemical potentials differ from those at constant density by terms of order $C^2/E_{\alpha\beta}$, where $C$ stands for the magnitude of the coupling parameters $C_{\alpha \para}$ and $C_{\alpha \perp}$, and $E_{\alpha\beta}$ for a typical coefficient in Eq.\ (\ref{Edens}).  This difference is of order $c_{ij}$ times a small quantity, the ratio of a surface (or Coulomb) energy density to a bulk energy density, and it is a good approximation to ignore it.  Thus we may calculate the properties of polycrystalline pasta phases by applying  the formalism developed for conventional solids.  Since it is a good approximation to treat the densities as constant, the ``bulk modulus'' to be used is that associated only with the strains but in the absence of density changes, namely $K^u$, not the thermodynamic bulk modulus, which also contains contributions from density changes.       

\subsection{Voigt and Reuss bounds}
Historically, the first approach to calculating effective elastic constants of polycrystals was that of Voigt \cite{Voigt}, who assumed that the \textit{strain} in every crystallite was the same.  As Hill demonstrated, this provides \textit{upper} bounds on the effective elastic constants of the polycrystal \cite{Hill}.
The general expression for the Voigt bound for the shear modulus is
\bea
 \mu_V=  \hspace {22em} \nonumber  \\  \frac1{15}\left[(c_{11} + c_{22} + c_{33}) - (c_{12} + c_{13} + c_{23}) + 3(c_{44} + c_{55} + c_{66})\right],\nonumber \\
 \hspace {-20em}
\eea
which for a crystal with hexagonal symmetry reduces to 
\bea
\mu_V =\frac{1}{5}\left(G_V^{\rm uni} +2c_{44} +2c_{66}\right),
\label{muV}
\eea
where
\be
G_V^{\rm uni}= \frac13 \left(\frac{c_{11}+c_{12}}{2} + c_{33}-2c_{13}\right)
\ee
is an effective elastic constant associated with a ``uniaxial shear stress'' with $(u_{xx}, u_{yy}, u_{zz})=(1,1,-2)/\sqrt 6$.  For more details, see Ref.~\cite{Berryman2004, Berryman2005}.
 
The general structure of Eq.~(\ref{muV}) and a number of later equations for shear elastic constants may be understood by thinking in terms of spherical tensors, rather than Cartesian ones.  Because a hexagonal crystal is elastically isotropic under rotations about its symmetry axis, the eigenfunctions of the elastic constant tensor may be classified according to their symmetry under rotations about that axis. Spherical tenors are specified by their degree $\ell$ and their order $m$, and for a symmetrical second rank tensor, for $m=\pm2$ or $\pm 1$, the only possible value of $\ell$ is 2.  Thus the eigenvalues of the elastic tensor are $c_{44}$ for $m=\pm 1$ and $c_{66}$ for $m=\pm2$.  For $m=0$, the eigenfunctions are superpositions of $\ell=0$ and $\ell=2$ contributions, which accounts for the more complicated form of the first term in Eq.~(\ref{muV}).  The factors of two in the second two terms come from the two possibilities for the sign of $m$ when $m\neq 0$,  and the factor of $1/5$ comes because the quantity is the average over the 5 different values of $m$: $0$, $\pm 1$ and $\pm 2$.     

For $K^u$, the Voigt bound is, in general,
\be
K^u_V=  \frac19\left(c_{11} + c_{22} + c_{33} + 2c_{12} + 2c_{13} + 2c_{23}\right)   ,
\ee
which for hexagonal symmetry reduces to
\be
K^u_V=\frac19[2(c_{11}+c_{12})+c_{33}+4c_{13}].
\ee

In the Reuss approximation, it is assumed that the \textit{stress} is constant throughout the medium, and Hill showed that this approximation gives \textit{lower} bounds on $\mu$ and $K^u$ \cite{Hill}.
The expression for the Reuss bound for the shear modulus for hexagonal symmetry is
\be
\frac1{\mu_R} =\frac15\left(\frac1{G_R^{\rm uni}} +\frac2{c_{44}} +\frac2{c_{66}}\right).
\ee
Here the quantity
\be
G_R^{\rm uni}= \frac{c_{33}(c_{11}+c_{12})-2c_{13}^2}{6K^u_V}
\ee
is an effective elastic constant associated with a ``uniaxial shear strain'' with $(\sigma_{xx}, \sigma_{yy}, \sigma_{zz})=(1,1,-2)/\sqrt 6$. 
This shows that the Reuss average is zero for unmodulated lasagna (both $c_{44}$ and $c_{66}$ vanish) and unmodulated spaghetti ($c_{44}$ is zero).
One can express $\mu_R$ explicitly in terms of elastic constants but the resulting expression is lengthy and not physically illuminating \cite[Eq.~(25)]{Kube_deJong}. 
The Reuss bound for $K^u$ is given by (see, e.g., Ref.\ \cite[Eq.~(24)]{Kube_deJong})
\be
K^u_{R}=\frac{c_{33}\left(c_{11}+c_{12}\right)-2 c_{13}^{2}}{c_{11}+c_{12}-4 c_{13}+2 c_{33}}.
\label{KR}
\ee
This also vanishes for unmodulated pasta, since $c_{11}, \, c_{12}$ and $c_{13}$ vanish for uniform lasagna and $c_{33}$ and $c_{13}$ vanish for uniform spaghetti, and therefore modulations are important for giving nonzero lower bounds for $\mu$ and $K^u$.  Generally the modulations of the pasta are small compared with those that give rise to the basic structure of the phases.  We shall consider $c_{33}$ for lasagna and $c_{11}$, $c_{12}$, and $c_{66}$ for spaghetti to be ``large'' and all other elastic constants to be ``small''.    One then finds to leading order in the small components that, for lasagna,
\be
\mu_V\simeq\frac{c_{33}}{15}, 
\ee
\be
K^u_V\simeq   \frac{c_{33}}{9}  ,
\ee
\bea
\mu_R\simeq   5\left(  \frac{2}{3(c_{11}+c_{12})}  +\frac{2}{c_{44}} +\frac{4}{c_{11}-c_{12}}  \right)^{-1}\\
= \frac{15(c_{11}^2-c_{12}^2)c_{44}}{6  (c_{11}^2-c_{12}^2)+c_{44}(10 c_{11}+14c_{12})},
\eea
and
\be
K^u_R\simeq \frac{c_{11}+c_{12}}{2},
\ee
and, for spaghetti,
\be
\mu_V\simeq  \frac{1}{30} (7c_{11}-5c_{12}),
\ee
\be
K^u_V\simeq  \frac29(c_{11}+c_{12})   ,
\ee
\bea
\mu_R\simeq \frac{15}{2}\left(\frac{1}{c_{33}}+\frac{3}{c_{44}}   \right)^{-1}\\
= \frac{15}{2}\frac{c_{33}c_{44}}{(3c_{33}+c_{44})} ,  
\eea
and
\be
K^u_R\simeq c_{33}.
\ee
The Reuss lower bounds are thus proportional to the small components of the elastic constant matrix but they are not analytic functions of the small components  in the limit when the small components tend to zero. 
The large difference between the Voigt and Reuss bounds points to the need for a better approximation.

\subsection{Self-consistent approach}

The basic idea in the self-consistent method is that one considers a crystalline inclusion embedded in an otherwise homogeneous medium with bulk modulus $K^u$ and shear modulus $\mu$.  There are a number of ways to arrive at the self-consistent formalism.  One is to consider applying a strain to the system and then asking how the strain is altered by the presence of the inclusion.  The self-consistent values of the effective elastic constants are determined by the condition that, on averaging over possible orientations of the inclusion, the change in the strain is zero.  A related way of arriving at the self-consistent effective elastic moduli it to imagine a transverse or longitudinal wave propagating in the homogeneous medium and calculate how the wave is scattered by the inclusion.  The moduli $K^u$ and $\mu$ are then calculated by demanding that there be no scattering on average for random possible orientations of the inclusion \cite{GubernatisKrumhansl}.  Another route to the results is to calculate improved upper and lower bounds on the effective elastic constants \cite{Kube_deJong}.
 For a crystal with hexagonal symmetry, the effective shear constant $\mu$ and bulk modulus $K^u$ are given in the self-consistent theory by the solution of the pair of coupled equations \cite{Berryman2005}
\bea
\frac{1}{\mu+\zeta}=\frac{1}{5}\left[\frac{1-\alpha\left(K^u_{\mathrm{V}}-K^u\right)}{G^{\mathrm{uni}}_{\mathrm{V}}+\zeta}+\frac{2}{c_{44}+\zeta}+\frac{2}{c_{66}+\zeta}\right]
\label{musc}
\eea
and 
\bea
K^u=\frac{G^{\mathrm{uni}}_{\mathrm{V}} K^u_{\mathrm{R}}+\zeta K^u_{\mathrm{V}}}{G^{\mathrm{uni}}_{\mathrm{V}}+\zeta}.
\label{Ksc}
\eea
Here
\be
\zeta=\frac{\mu}{6}\left(\frac{9 K^u+8 \mu}{K^u+2 \mu}\right),
\ee
and 
\be
\alpha=-\frac{1}{K^u+{4\mu}/3}.
\ee

For laboratory solids and for most geological minerals, the Voigt and Reuss bounds are relatively close.  However, as we have seen, the Reuss lower bounds on the effective bulk and shear moduli of pasta phases without modulations are zero, so the bounds are not very useful in pinning down these moduli.  As we shall describe below, the self-consistent formalism also leads to vanishing effective elastic constants.  We thus consider the effects of the spatial modulations of the pasta.  
To implement the self-consistent approach, we have adopted two methods.  The first is to solve the equations (\ref{musc}) and (\ref{Ksc}) directly.  These equations are polynomials in $\mu$ and $K^u$ and have many solutions, but only one of them is physically meaningful.  The other method is to employ the computer algorithm given in Ref.\ \cite{Kube_deJong}.  This calculates successively better upper and lower bounds on $\mu$ and $K^u$ for arbitrary crystal structures.   In high order the two bounds converge to the self-consistent value.  We have confirmed numerically that the two procedures lead to identical results.

The first important result of the self-consistent approach is that, in the limit of uniform lasagna and uniform spaghetti, the self-consistent values of $\mu$ and $K^u$ all vanish.  Thus, in these cases the self-consistent elastic constants coincide with the Reuss lower bound.    The approximation suggested by Hill \cite{Hill}, that one take the arithmetic average of the Voigt and Reuss bounds, while it generally gives sensible results for laboratory solids, is thus very misleading for the pasta phases. 

\begin{figure}
\includegraphics[width=3.5in]{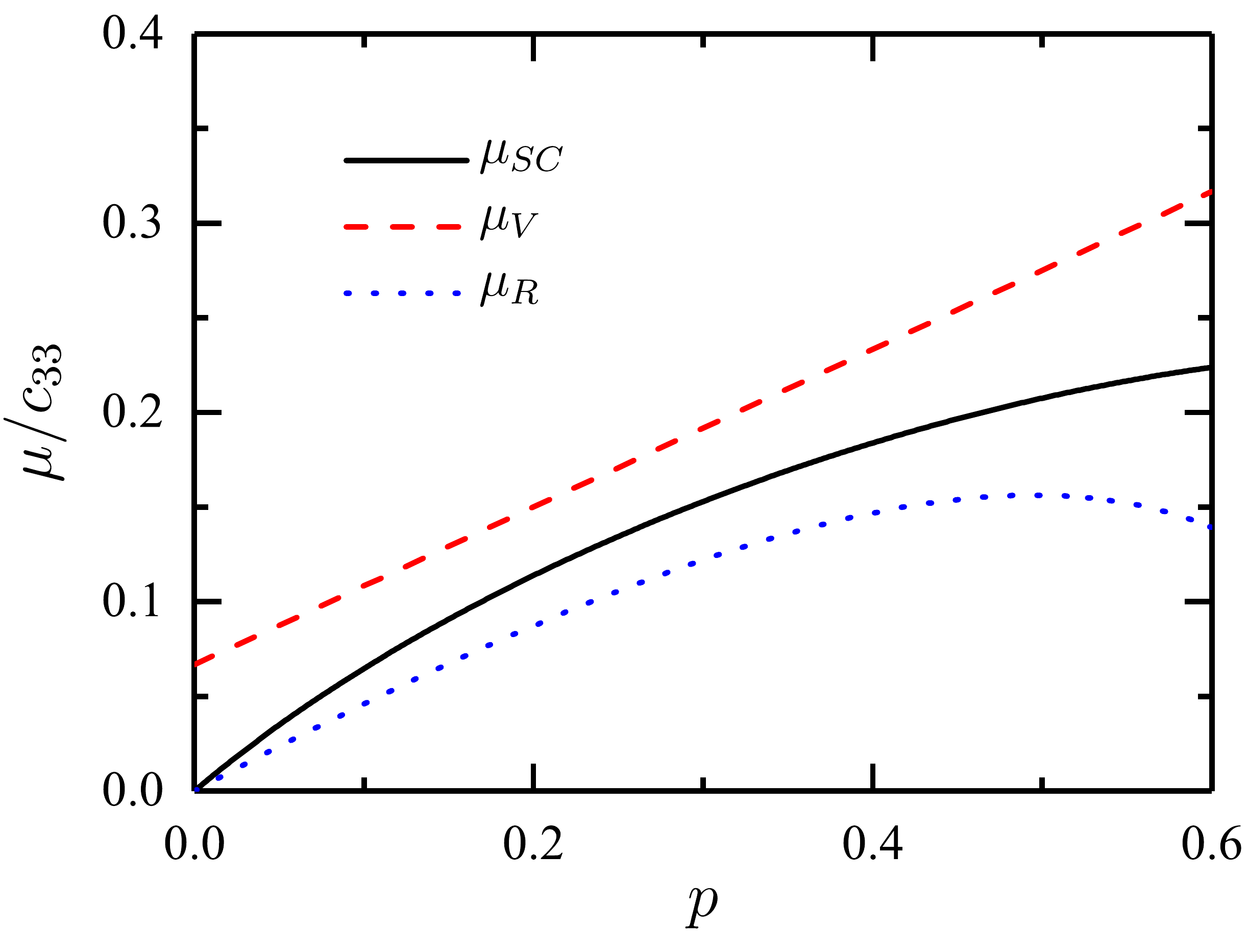}
\caption{Effective shear elastic constant of polycrystalline lasagna divided by $c_{33}$ as a function of $p$, which is a measure of the smallness of the elastic constants that vanish for unmodulated lasagna..   For details, see the text. The self-consistent result is shown as the black solid line, the Voigt upper bound as the red dashed line, and the Reuss lower bound as the blue dotted line.}
\label{lasagnaplotshear}
\end{figure}

\begin{figure}
\includegraphics[width=3.5in]{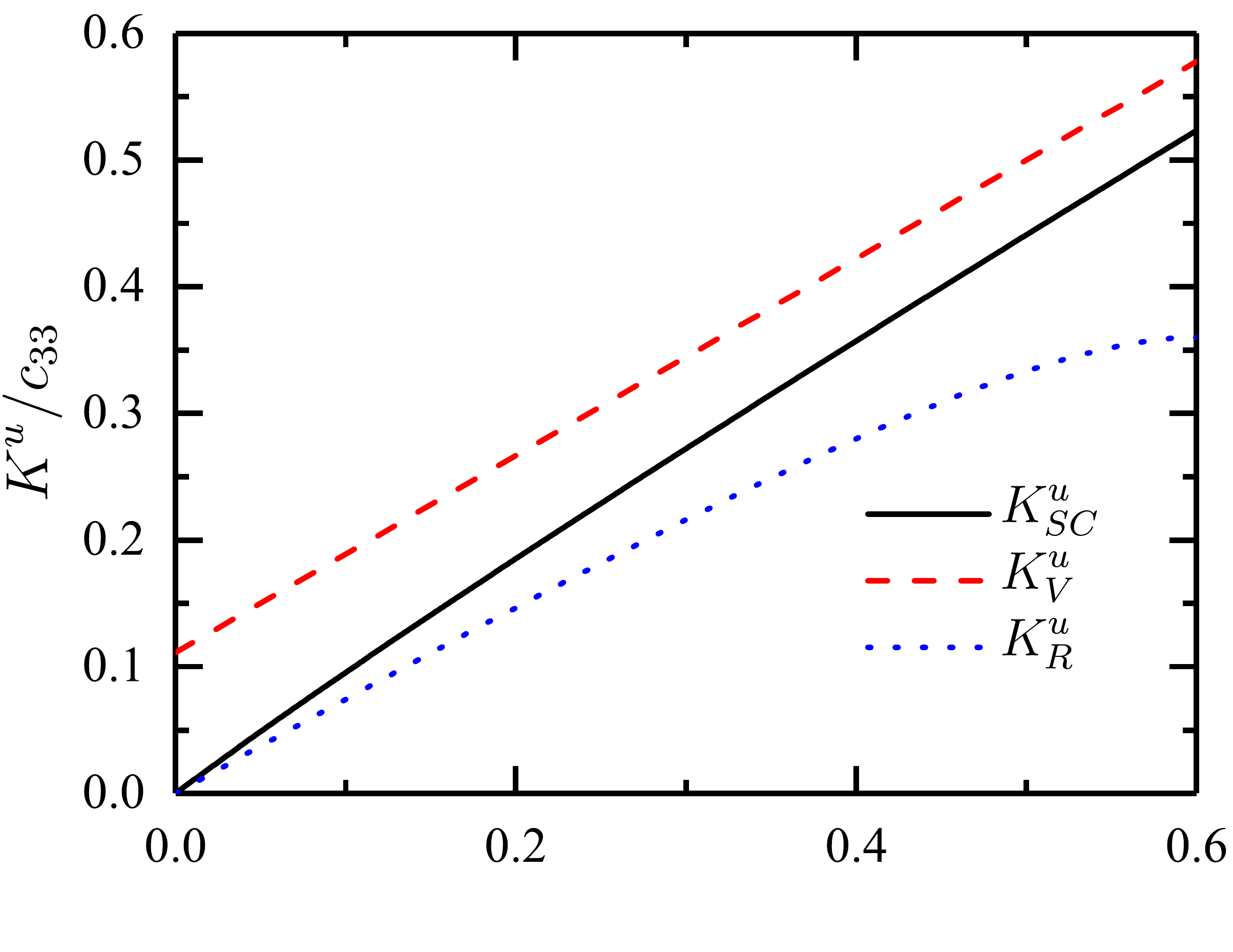}
\caption{Effective bulk elastic constant $K^u$ of polycrystalline lasagna divided by $c_{33}$ as a function of $p$.  For details, see the caption to Fig.\ \ref{lasagnaplotshear} and the text.}
\label{lasagnaplotbulk}
\end{figure}

These results show that in the self-consistent formalism, the ``small'' elastic constants must be included in order to obtain nonzero elastic constants for the polycrystal.  These elastic constants have not yet been calculated, but in the simulations of lasagna in reference \cite{Caplan} it was found that disconnected plates have near zero shear modulus for sliding of sheets on one another, i.e., $c_{44}\approx 0$.   We now present calculations in the self-consistent formalism for lasagna for a range of choices of the small elastic constants.   For stability, the elastic constants must obey the conditions \cite{MouhatCoudert}
\be
c_{11} > |c_{12}|, 
\label{stab1}
\ee
\be
c_{44}>0,
\label{stab2}
\ee
and
\be 
2c_{13}^2 < c_{33}(c_{11} + c_{12}).
\label{stab3}
\ee
We introduce a parameter $p>0$ which is a measure of the size of the small elastic constants compared with the large one, $c_{33}$.   We choose all other elastic constants than $c_{33}$ to be $p c_{33}$, except $c_{12}=pc_{33}/2$ (and consequently $c_{66}=pc_{33}/4$), thereby satisfying the conditions (\ref{stab1}) and (\ref{stab2}). Condition (\ref{stab3}) is satisfied for $p<3/4$. We plot the result of the self-consistent theory and the Voigt and Reuss bounds for the effective shear modulus of lasagna as a function of $p$ in Fig.\ \ref{lasagnaplotshear} and the corresponding results for $K^u$  are shown in Fig.\ \ref{lasagnaplotbulk}.

\begin{figure}
\includegraphics[width=3.5in]{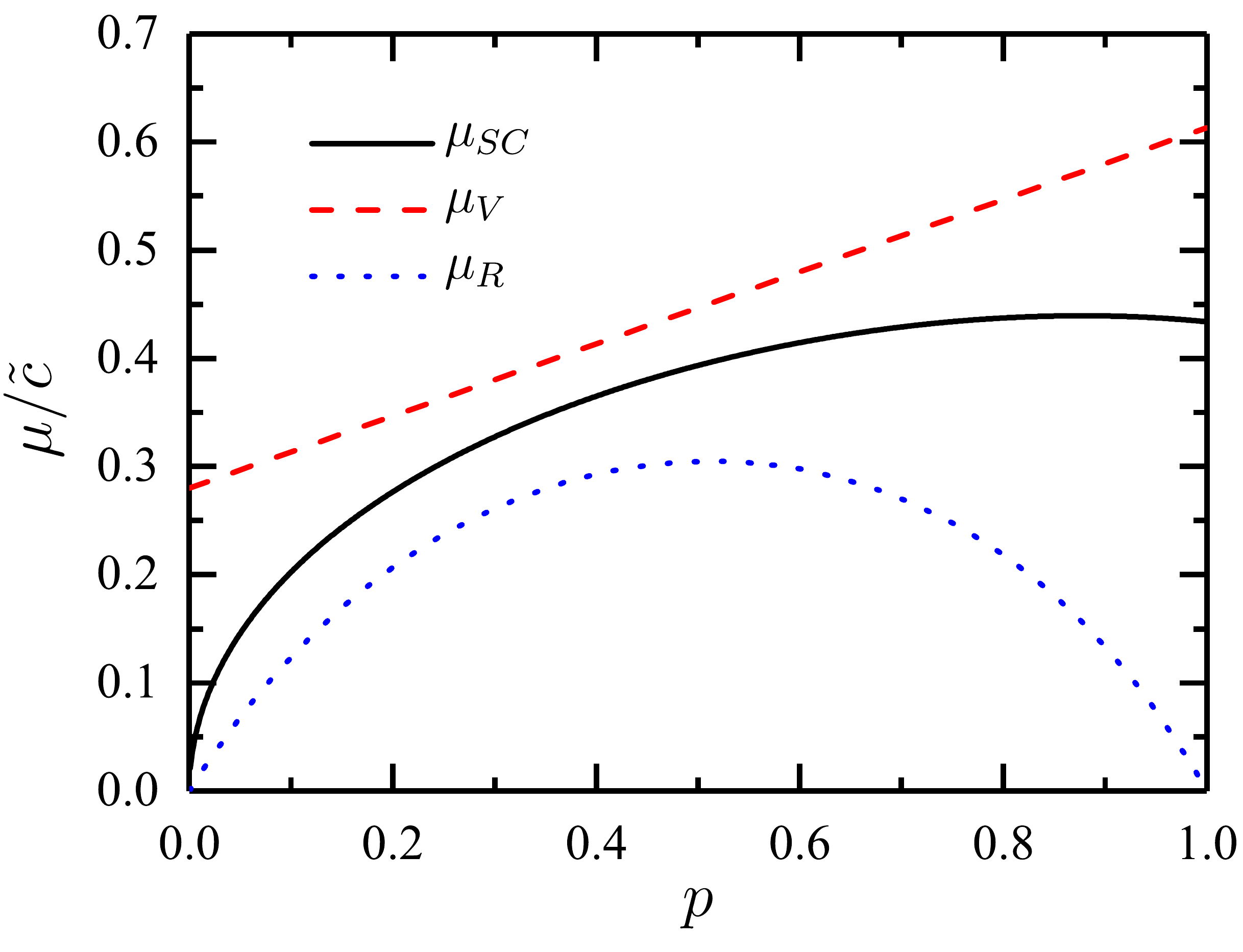}
\caption{Effective shear elastic constant of polycrystalline spaghetti divided by $\tilde c=(c_{11}+c_{12})/2$.  For details, see the caption to Fig.\ \ref{lasagnaplotshear} and the text.   }
\label{spaghettiplotshear}
\end{figure}

As we have previously stated, for small $p$ the effective shear modulus and $K^u$ tend to zero, and are linear in $p$.  Offhand, one might expect the modulations of the pasta structures to have amplitudes in the range 0.1--0.3 times the amplitudes of the modulations responsible for the basic pasta elements and the ``small'' elastic constants, which vary as the square of the amplitudes, to be in the range $10^{-2}$ to  $10^{-1}$ times $c_{33}$ for lasagna or $c_{11}$ for spaghetti, corresponding to $p\sim 10^{-2}$--$10^{-1}$.  For such values of $p$ the self-consistent elastic constants are considerably closer to the Reuss lower bound than to the Voigt upper bound and, as a first approximation one may use the Reuss bound.

It is important to bear in mind that the effective bulk modulus $K^u$ calculated here is that for \textit{fixed particle densities}.  Consequently, its vanishing does not imply that matter is on the verge of an instability to collapse.

\begin{figure}
\includegraphics[width=3.5in]{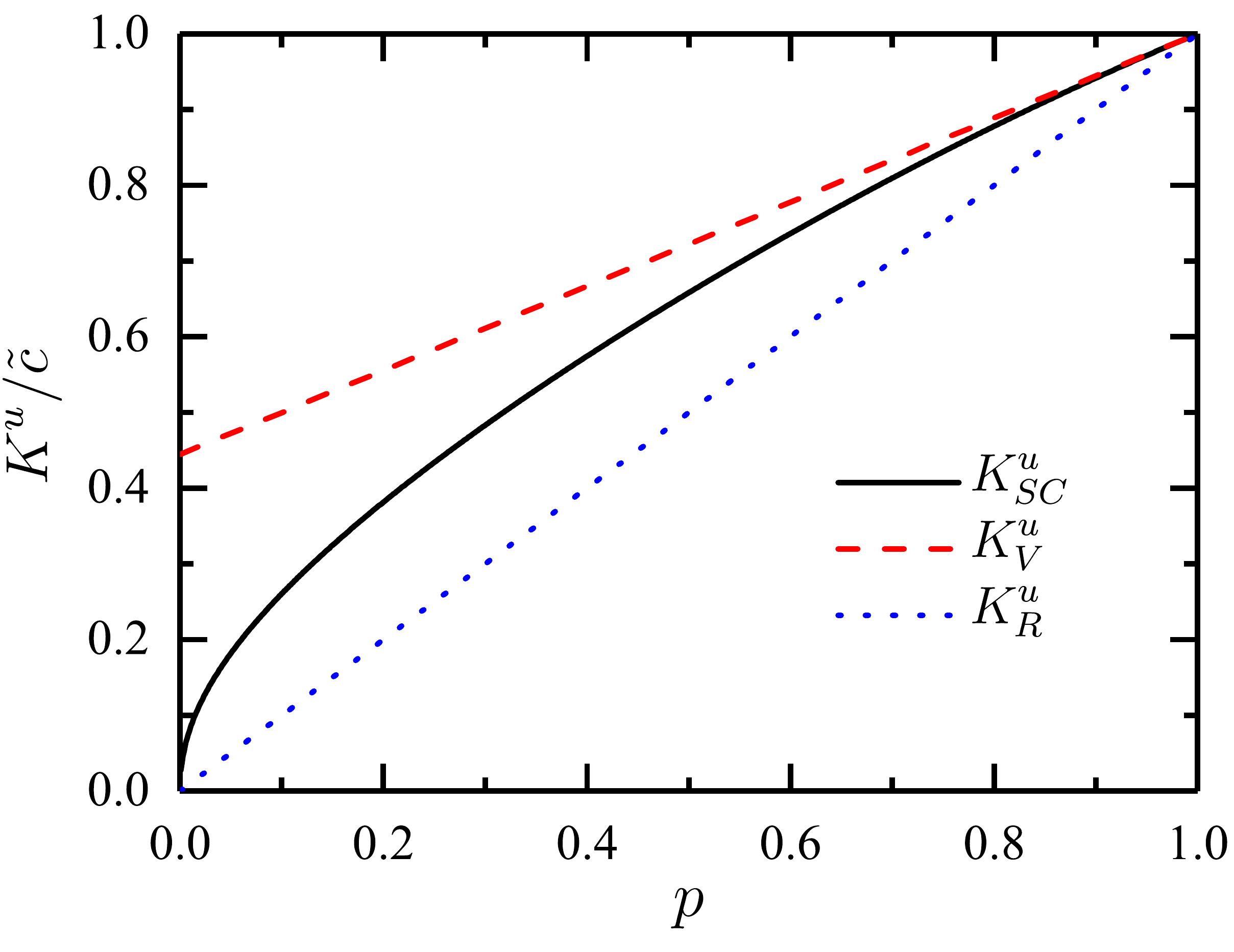}
\caption{Effective bulk elastic constant $K^u$ of polycrystalline spaghetti divided by $\tilde c=(c_{11}+c_{12})/2$.   For details, see the caption to Fig.\ \ref{lasagnaplotshear} and the text.   }
\label{spaghettiplotbulk}
\end{figure}

We turn now to spaghetti, which, without spatial modulations, has two independent nonzero elastic constants, $c_{11}$ and $c_{12}$ or, equivalently, $(c_{11}+c_{12})/2$ and   $c_{66}=(c_{11}-c_{12})/2$.    According to the calculations of Refs.\ \cite{pasta, WilliamsKoonin, Oyamatsu}, the spaghetti phase is expected to be stable in a range of filling factors $0.15\lesssim u\lesssim 0.35$, and in the middle of this range, $u\approx 0.25$, one finds that $g(u)\approx 0.75$, which implies that  $   (c_{11}-c_{12})/(c_{11}+c_{12})\approx 0.5$.  We have carried out calculations of the effective elastic constants for polycrystals for this value of the ratio, and have taken the other elastic constants ($c_{13}$, $c_{33}$, and $c_{44}$) to be $p (c_{11}+c_{12})/2$.   Results are plotted in Figs.\ \ref{spaghettiplotshear} and \ref{spaghettiplotbulk}.  The range of values of $p$ used is larger than is expected to occur in practice in order to bring out the interesting behavior at $p=1$.  The Reuss bound  for $\mu$ vanishes for $p=1$, since $G_R^{\rm uni}$ vanishes for that case: the lattice becomes unstable since condition (\ref{stab3}) is no longer satisfied.  The fact that the Reuss average for $K^u$ is a linear function of $p$ is also special, since generally it is not.  The reason for this can be traced to the fact that, by chance, the elastic constants satisfy the condition $c_{13}=c_{33}$ and therefore, from Eq,\ (\ref{KR}), $K^u_R=p(c_{11}+c_{12})/2$.

A striking feature of the self-consistent results is that $\mu$ and $K^u$ vary as $p^{1/2}$ for small $p$, in contrast to what was found for lasagna.  A similar effect is found in an effective medium approach  to calculate the electrical conductivity of polycrystals made up of crystallites of a uniaxial crystal, \cite[Sec.\ IIIB]{Stroud}.

\begin{figure}
\includegraphics[width=3.5in]{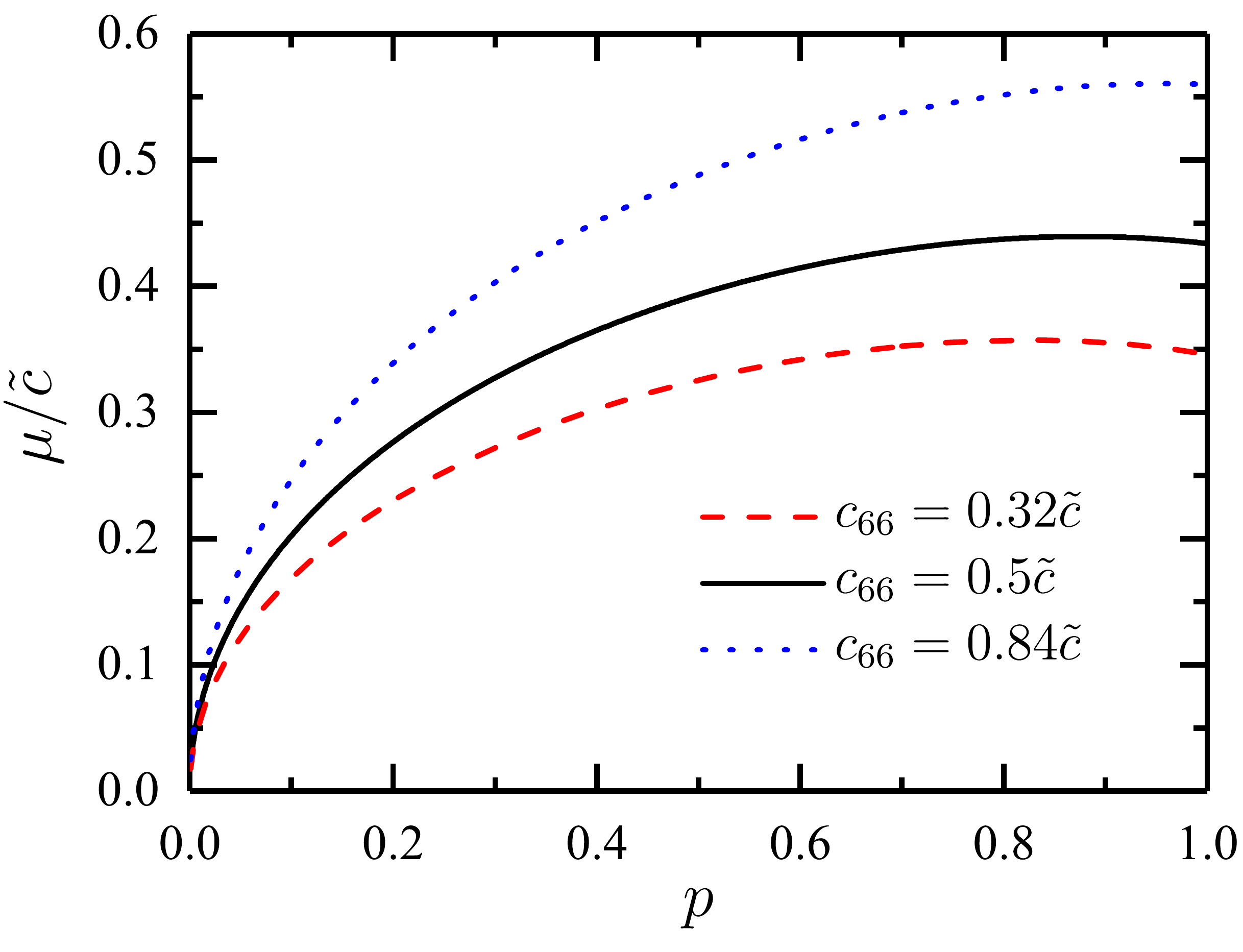}
\caption{Effective shear elastic constant of polycrystalline spaghetti divided by $\tilde c$ as a function of $p$ for various values of the ratio $\lambda=c_{66}/\tilde c$.    For details, see the text.  }
\label{Shear-C66}
\end{figure}

\begin{figure}
\includegraphics[width=3.5in]{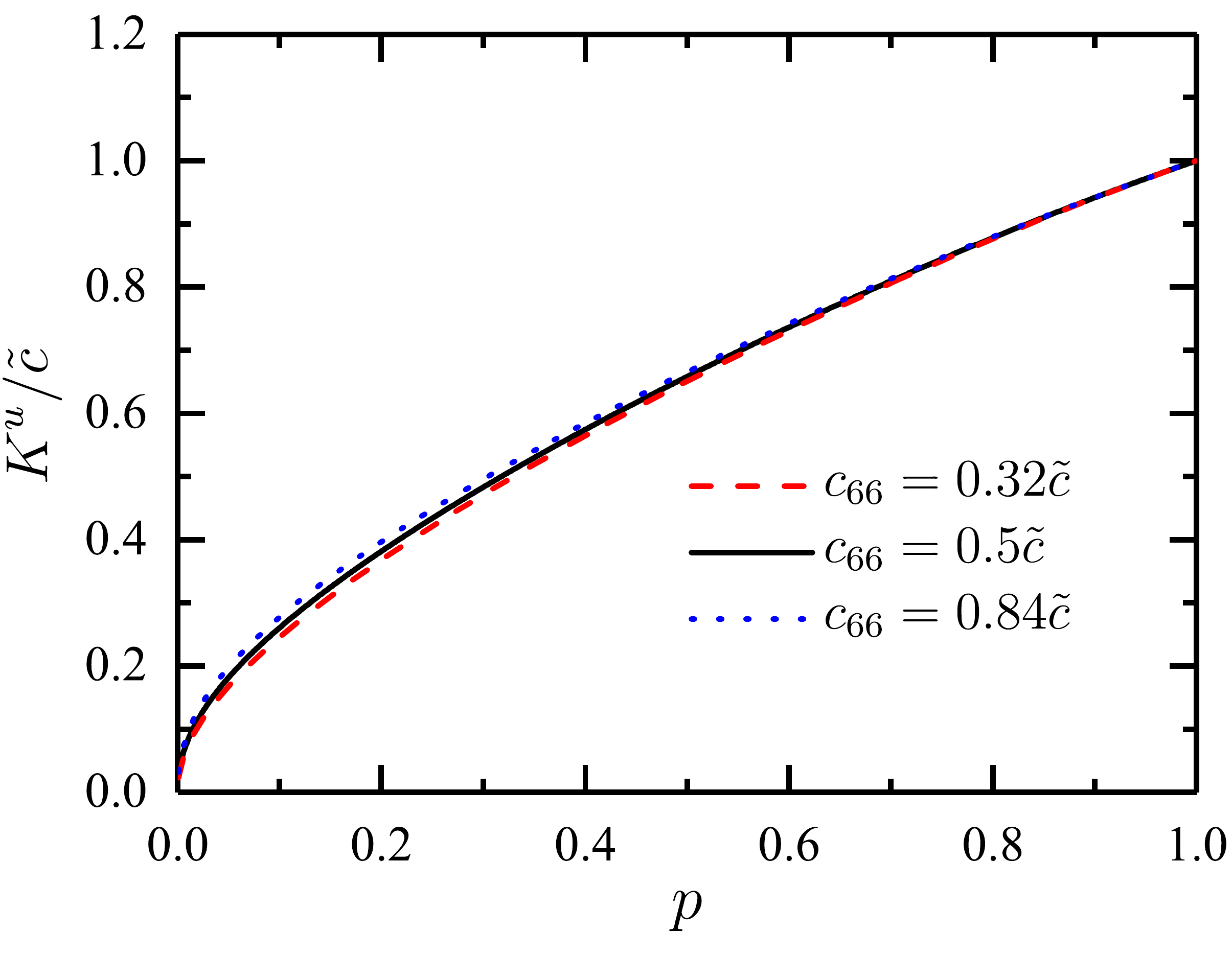}
\caption{Effective bulk elastic constant $K^u$ of polycrystalline spaghetti divided by $\tilde c$ as a function of $p$ for various values of the ratio $\lambda=c_{66}/\tilde c$.    For details, see the text.   }
\label{Bulk-C66}
\end{figure}

We have also calculated $\mu$ and $K^u$ for other values of the ratio $\lambda=(c_{11}-c_{12})/(c_{11}+c_{12})=c_{66}/\tilde c$ and the results are plotted in Figs.\ \ref{Shear-C66} and \ref{Bulk-C66}.  Besides the value $\lambda=0.5$ that we used in the earlier calculations, we have also used the values $\lambda=0.32$ appropriate for a filling factor of 0.15
and 0.84  for a filling factor of 0.35.  The other elastic constants are taken to have the form used in the calculations described earlier.  For fixed $\tilde c$, $\mu$ is much more sensitive to $\lambda$ than $K^u$ is.  This is natural because $K^u$, which is a measure of the rigidity of the system to a homologous compression, is little affected by shear elastic constants such as $c_{66}$.


\section{Summary and concluding remarks}
\label{summary}

In this paper, we have made estimates of the elastic constants of pasta phases without spatial modulations.  We have also calculated the effective shear modulus and $K^u$ of these phases when they are polycrystalline or, more generally, do not exhibit long-range spatial order.   We have found in the self-consistent approach to calculating effective elastic moduli, that spatial modulation of the pasta elements plays a crucial role: for lasagna $\mu$ and $K^u$ are proportional to the parameter $p$, which is a measure of the size of the small components of the elastic constants, while for spaghetti, they are proportional to $p^{1/2}$.

An important finding is that the use of the Voigt upper bound to predict elastic properties of polycrystals, which has been the common practice in the astrophysical literature until recently, gives qualitatively incorrect results for the pasta phases. 

The elastic properties calculated in this paper are one of the  necessary ingredients in calculations of the frequencies of collective modes.   Another is the superfluid density tensor for pasta phases without long-range order, which can be calculated by methods analogous to those described here, but this has not yet been done. 

In this paper we have not taken magnetic fields into account.  Some effects of a magnetic field in the pasta phases have been discussed in Ref.\ \cite{Kobyakov2018}.  For normal, i.e., non-superconducting, matter, magnetic contributions to the stress tensor have a magnitude of order $B^2/8\pi$, where $B$ is the magnetic flux density.  Even for $B\sim 10^{15}$ gauss, a typical value for a magnetar, the magnetic stresses are of order $10^{29}$ erg cm$^{-3}$ which is small compared with the estimates we have made for the elastic constants.   If matter is superconducting, components of the magnetic stress tensor are $\sim B H/4\pi$, where $H$ is the magnetic field \cite{EassonCJP}.   For superconducting matter, $H$ can be very much greater than $B$.  For $B<10^{15}$ gauss, estimates of $H$ are less than $10^{15}$ oersted, and consequently the components of the stress tensor are $\lesssim 10^{29}$ erg cm$^{-3}$ even when matter is superconducting.   For $B\sim 10^{12}$ gauss, which is typical of pulsars, the effects of a magnetic field are considerably smaller.   Magnetic fields could increase the rigidity of matter, an effect that would be equivalent to increasing the stresses described by the ``small'' elastic constants.   In regions where superconductivity is on the verge of disappearing, magnetic forces could be significant because gradients of $H$, which for the low flux line densities of interest is close to the lower critical field $H_{c1}$, can be large. A more detailed investigation of these effects is warranted.

As a consequence of the anisotropy of the pasta phases, for a given magnetic flux density the magnetic field in the pasta phases is anisotropic, and this will induce a torque on the matter which tends to align $\bf B$ with the direction in the matter with the lowest magnitude of $\bf H$ for the given value of $B$.  This would result in the magnetic field tending to lie in the plane of the lasagna sheets and in the direction of the spaghetti strands

There are a number of other open questions.  What are the conditions for  the pasta phases to have a lower energy than uniform matter?  To elucidate this requires the use of modern developments in the theory of nucleon--nucleon interactions to calculate properties of the pasta phases.  How large are the spatial modulations of the pasta phases?  Here it would be useful to have a better analytical understanding of the origin of the modulations.  In addition, in order to test the predictions of the self-consistent theory, it would be valuable to find terrestrial analogs of the modulated pasta phases and also to perform numerical simulations of systems that contain in the computational cell a larger number of crystallites than was the case in Ref.\ \cite{Caplan}.  \\

 \section*{Acknowledgments} We are grateful to Matt Caplan for helpful correspondence on his simulations of the pasta phases and to J\o rgen Randrup for valuable comments on applications of molecular dynamics methods to degenerate systems.

\end{document}